\documentclass[manuscript]{aastex62}

\usepackage{xcolor}
\usepackage{booktabs}
\usepackage{hyperref}
\graphicspath{{./}{figures/}}

\submitjournal{ApJ}

\shorttitle{Energetic Electron Path Lengths}
\shortauthors{Zhao et al.}

\begin{document}

\title{Statistical analysis of interplanetary magnetic field path lengths from solar energetic electron events observed by \it{WIND}}

\author[0000-0003-3936-5288]{Lulu Zhao}
\affil{Department of Aerospace, Physics and Space Sciences, Florida Institute of Technology, Melbourne, FL, 32901, USA, lzhao@fit.edu}

\author[0000-0003-4695-8866]{Gang Li}
\affiliation{Department of Space Science,University of Alabama in Huntsville, Huntsville, AL, 35899, USA, gangli.uah@gmail.com}

\author{Ming Zhang}
\affil{Department of Aerospace, Physics and Space Sciences, Florida Institute of Technology, Melbourne, FL, 32901, USA}

\author{Linghua Wang}
\affil{School of Earth and Space Sciences,
Peking University, Beijing, 100871, China}

\author{Ashraf Moradi}
\affil{Department of Space Science,University of Alabama in Huntsville, Huntsville, AL, 35899, USA}

\author{Frederic Effenberger}
\affil{Helmholtz Centre Potsdam, GFZ German Research Centre for Geosciences, Potsdam, 14473, Germany}
\affil{Bay Area Environmental Reserach Institute, Petaluma, CA, 94952, USA}



\begin{abstract}
We calculate the interplanetary magnetic field path lengths traveled by electrons in solar electron events detected by the WIND 3DP instrument from $1994$ to $2016$.
The velocity dispersion analysis method is applied for electrons at energies of $\sim$ $27$ keV to $310$ keV.
Previous velocity dispersion analyses employ the onset times, which are often affected by instrumental effects and the pre-existing background flux, leading to large uncertainties. 
We propose a new method here. Instead of using the peak or onset time, we apply the velocity dispersion analysis to the times that correspond to the rising phase of the fluxes that are a fraction, $\eta$, of the peak flux. 
{We perform statistical analysis on selected events whose calculated path lengths have uncertainties smaller than $0.1$ AU. The mean and standard deviation, ($\mu$, $\sigma$), of the calculated path lengths corresponding to $\eta=$ $3/4$, $1/2$, and $1/3$ of the peak flux is ($1.17$ AU, $0.17$ AU), ($1.11$ AU, $0.14$ AU), and ($1.06$ AU, $0.15$ AU).
The distribution of the calculated path lengths is also well fitted by a Gaussian distribution for the $\eta=3/4$ and $1/3$ cases.}
These results suggest that in these electron events the interplanetary magnetic field topology is close to the nominal Parker spiral with little field line meandering. Our results have important implications for particles' perpendicular diffusion. 
\end{abstract}

\section{Introduction}\label{sec:intro}
Solar electron events are common phenomenon observed in interplanetary space.
They are often observed in the energy range of $<1$ keV to $>300$ keV with an occurrence rate near the earth of $\sim190$ events per year during solar maximum and $\sim10$ per year during solar minimum \citep{Wang2012}. 
Energetic electrons are accelerated either in solar flares or in the vicinity of shocks driven by coronal mass ejections (CMEs).
The escaping accelerated electrons at energies of $2$ to $\gtrsim10$ keV \citep{Lin1985} excite type III radio bursts when propagating through the plasma of the solar corona and interplanetary space.
$98.75\%$ of the observed solar electron events in solar cycle $23$ are reported by \citet{Wang2012} to be accompanied by type III radio bursts. 
\citet{Wang2006} examined three scatter-free impulsive electron events and found that $0.4$-$10$ keV electrons were injected $9.1\pm4.7$ minutes before the type III bursts, while electrons at energies of $13$-$300$ keV were injected $7.6\pm1.3$ minutes after the type III bursts.
The delayed injection of high-energy electrons after the type III radio burst has been reported by \citet{Krucker1999}, \citet{Haggerty2001}, \citet{Haggerty2002}, \citet{Haggerty2003}, \citet{Cane2003a}, \citet{Cane2003}, \citet{Simnett2002}, \citet{Maia2004}, and \citet{Klein2005}. 
And the reason was suggested to be either high energy electrons are released at a later time (due to propagation of large-scale coronal transient \citep{Krucker1999}, shock waves \citep{Simnett2002}, or coronal magnetic restructuring after CMEs \citep{Maia2004, Klein2005}), or that the high energy electrons experience more interplanetary scattering \citep{Cane2003a, Cane2003}. 
By calculating the extended path length traveled by high-energy electrons compared to low-energy electrons, due to the interplanetary scattering, \citet{Wang2011} suggested that such scattering is too small to explain the observed delays.

Electrons of tens of keV are great tracers of the interplanetary magnetic field topology because of their fast speed and small gyro radius, in the order of $10^{-6}$ AU near the earth.
By measuring the direction to the centroid of the type III radio emission region and using the Helios model of the interplanetary plasma density \citep{Bougeret1984}, \citet{Reames1986} were able to map out the interplanetary magnetic field path traveled by electrons.
They found that in some events, the magnetic field paths depart from Parker spiral.
Assuming the release time of energetic electrons matches that of the type III radio burst in the high frequency, \citet{Larson1997} calculated the path length in a twisted flux-rope topology of magnetic clouds and found that the path length is $\sim3$ AU near the leading edge of the magnetic cloud and $\sim1.2$ AU near the center. 
Using the same method, \citet{Kahler2011, Kahler2011a} investigated the field line lengths in magnetic clouds and interplanetary coronal mass ejections. They found that there was no significant difference of the path length among magnetic clouds, ICMEs, and ambient solar wind. Moreover, the field line lengths at different parts of ICMEs do not show obvious differences.

Interplanetary magnetic field length has also been investigated using the observed magnetic turbulence power spectra as a function of resolution scale and turbulence condition.
\citet{Ragot2006} simulated the interplanetary magnetic field lines using $10^{15}$ different turbulent modes and computed the magnetic field lengths numerically. 
She found that, comparing to the case of a smooth field, a turbulent field line can be  $\sim 1.3$ times longer. 

Besides these attempts, another widely used method for determining the interplanetary magnetic field path length is the velocity dispersion analysis (VDA). By fitting the onset time of energetic electrons or ions at different energy channels, one can obtain the particle's release time at the sun and the path length simultaneously.
The validity of the VDA is based on several assumptions including the simultaneous release of particles of different energies on the sun; the scatter-free propagation of the first arriving particles; and the accurate determination of onset time in in-situ observations.
\citet{Kahler2006} examined the $c/v$ onset plots of energetic electron events and found the fitted path length distributed between $0.15$ and $2.7$ AU. And out of $80$ events surveyed,  $64$ events have path length smaller than the minimum travel distance calculated from the instantaneous in-situ solar wind speeds. 
\citet{Kahler2006} attributed the reason for the unphysical short path lengths to be either the instrumental effects that produce early onset in lower energy, or the invalid assumption of impulsive and simultaneous injection, or the diffusive propagation.
The validity of VDA has also been analyzed numerically by \citet{Lintunen2004}, \citet{Saiz2005}, \citet{Laitinen2015}, and \citet{Wang2015}.

Clearly, the work of \citet{Kahler2006} indicated that not all events are suitable for VDA analysis: for example, events having very gradual rising time profiles may violate the assumption of energy-independent releasing assumption. Furthermore, both the onset times and  the peak times can be very hard to determined. These add further issues to the 
commonly used VDA.  To address these issues, we consider a new VDA analysis approach and use this approach to study the interplanetary magnetic field path length in selected solar electron events. 

As discussed above, to reduce the potential discrepancy of the release time between the low and high energy electrons, we focus on electrons at energies of $27-310$ keV. Furthermore, we choose events that have rapid rise and reasonably rapid decay phases. Such a time profile often indicates a scatter-free propagation and an impulsive, nearly symmetric injection at the sun \citep{Lin1974, Wang2006}.

\section{Event Database Description} \label{sec:data}
We use the electron measurements made by the 3-D Plasma and Energetic Particle (3DP) instrument (Silicon semiconductor telescope (SST) and electron electrostatic analyzers (EESA-L and EESA-H)) onboard WIND spacecraft \citep{Lin1995}.
The EESA-L and EESA-H detectors measure $\sim3$ eV to $30$ keV electrons with angular resolution of $22.5^\circ\times22.5^\circ$ and the SST measures $\sim 25$ to $400$ keV electrons with angular resolution of $22.5^\circ\times36^\circ$. 
Between December 1994 and October 2016, there are $1944$ solar energetic electron events identified with a flux increase of 2 times the standard deviation of the background flux above the background in three or more energy channels. The presence of a velocity dispersion with faster electrons arriving earlier than slower electrons is also required in selecting these electron events.
Among those $1944$ events, we select $882$ events that are detected by at least four energy channels by the SST detector, at energies of $27-310$ keV. 

Due to the instrumental effects of SST, a portion of incident high energy electrons could be scattered out of the detector and contaminate the low energy channels, which leads to the early detection of the lower energy particles \citep{Wang2006}.
\citet{Tan2013} compared the path lengths calculated with and without correcting the instrumental effect and suggested that the unphysical short path lengths calculated in \citet{Kahler2006} could be due to the contamination from the deposition energy loss.
Later, in stead of using the corrected onset times, \citet{Wang2016} used the peak times to perform the path length analysis.
Not only the contamination from higher energy channels is minimum at the peak time, the effect of the background superhalo electrons is also minimized.
However, since the peak is where the rate of flux change is zero, so deciding the correct peak time with small uncertainties can be hard. 
Using peak time to perform the VDA implicitly assumes that the peak of the injection profile of different energies are at the same time and the injection duration at the sun is short.
{Due to scattering, the peak of the time intensity profile  observed at 1 AU will be broader than that at the source. In the work of \citet{Wang2006}, to better obtain 
the peak time, events with short duration of injection are selected.  As 
we will see below, in the new method, we do not need to accurately determine
 the peak time. However, a short duration of injection often suggests a 
 rapid rising phase. And a rapid rising phase leads to a smaller uncertainties of the obtained path length. For this reason, we focus on electrons with higher energies ($27-310$ keV), which have relative shorter injection duration compared to low-energy electrons \citep{Wang2006, Wang2016}}.

Unlike the peak, which may be hard to identify, the rising phase in a majority of electron events (certainly impulsive ones and also some graduate ones) is very rapid and the shapes are similar at different energy channels. Noticing this feature, we develop a new VDA procedure to reduce the effect of a gradual decay phase on the determination of the peak times. 
We first find the peak value, $j_p$, in the time intensity profile of outward-propagating electrons.
Note that while peak time is hard to determine, peak flux determination 
is easy. Furthermore, our method does not require an accurate 
value of the peak flux (see below).

\begin{figure}[ht!]
\plotone{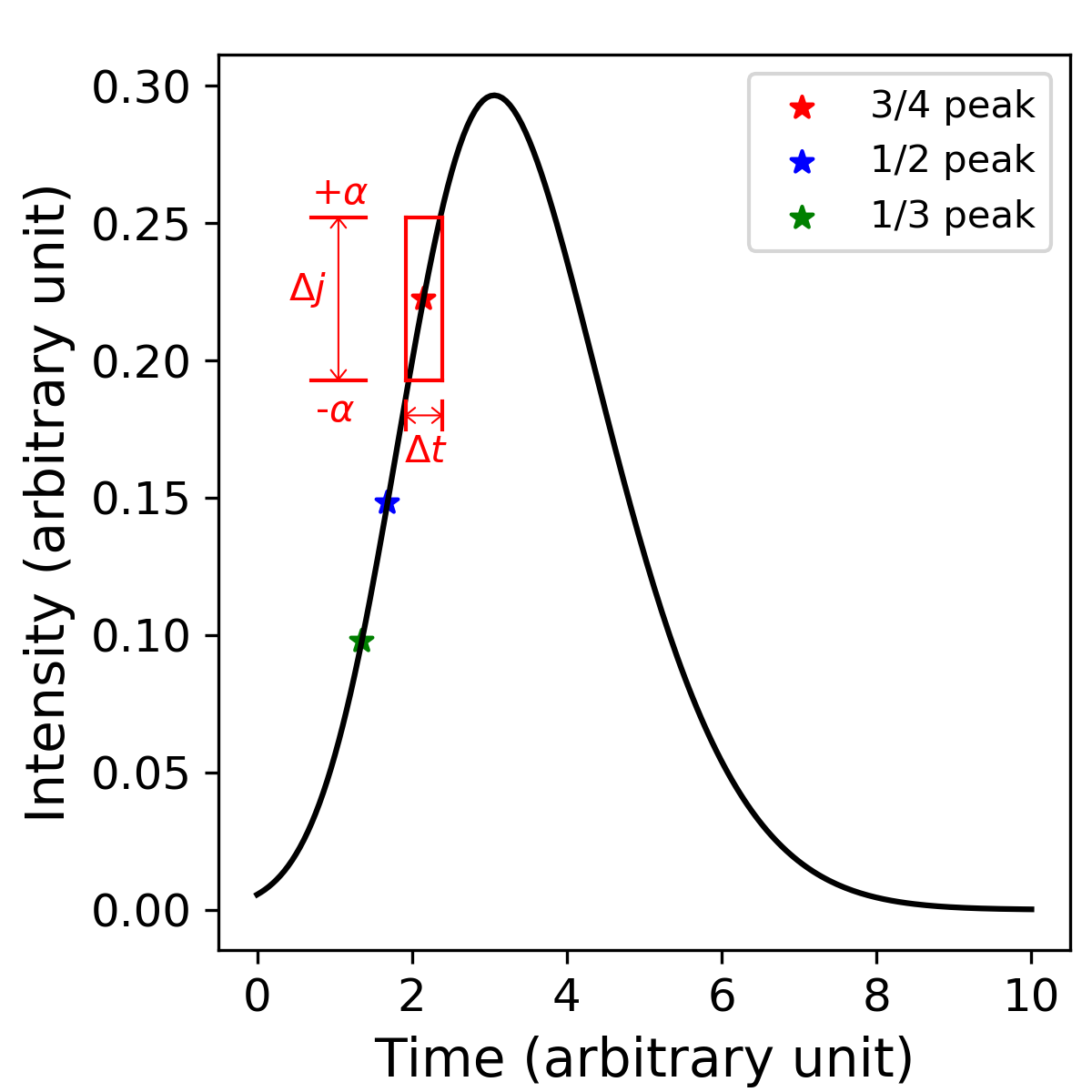}
\caption{Illustration of the FVDA method. The black curve is the time intensity profile observed. The red, blue, and green asterisk symbols on the curve represent the point whose intensity is $3/4$, $1/2$, and $1/3$ of the peak intensity. The red box demonstrate the method that we determine the uncertainty associated with the time when the intensity is $3/4$ of the peak value. The vertical boundary represents the $\pm\alpha$ of the flux and the horizontal boundary represents the corresponding time uncertainty.}\label{fig:method}
\end{figure}

We next find the times $t_\eta$ in the rising phase that satisfy 
\begin{equation}
    j(t_\eta)-j_b=\eta (j_p-j_b) 
\end{equation} 
with $\eta$ to be the fraction parameter. In the 
above, $j_b$ is the background flux. 
The background flux $j_b$ is taken to be the minimal flux in a $20\sim30$ minutes window before or after the event.
Because the rising phases are sharp, and the fact that 
the rising phases for different energy channels are similar, we can use  $t_{\eta}$'s for the corresponding VDA analysis.
We further assume  a $\Delta j = \alpha  j(t_\eta)$ to obtain the uncertainty of $t_\eta$ at corresponding $j(t_\eta)$ as illustrated in figure~\ref{fig:method}.
The red rectangular box represents the method that we determine the uncertainty of the time $t_\eta$, $\Delta t_\eta$. 
The vertical boundary represents the $\pm \alpha$ of the flux uncertainty region and the corresponding horizontal boundary represents the uncertainty of $t_\eta$.
The uncertainties, $\Delta t_\eta$, are used as weights in the Chi-square fitting method that we utilize to perform the linear regression. 
Therefore, different values of $\alpha$ may yield slightly different fitting results and their corresponding uncertainties. 
In this work, the value of $\alpha$ is chosen to be $7.5\%$, sufficiently large to account for the time resolution of the observations ($\sim 24 s$), the uncertainty in determining the peak flux, and the contamination due to electrons from higher energy channels. 

Note that the contamination from higher energy channels occur mainly at the onset so that it will lead to inaccurate identification of the onset times and a smaller path length. Indeed, as shown in \citet{Kahler2006}, in performing the traditional VDA analysis, one may obtain path lengths that are as small as $0.15$ AU due to the contamination effect. And the contamination from the high energy channel is event-based, therefore, it is hard to estimate the effect of contamination on path length in a quantitative way.  As an event progresses, the contamination becomes less significant since the fraction of particles within its original energy bin increases. 
In \citet{Li2013}, the time intensity profile before and after correcting the contamination from the higher energy channels is plotted and the contamination effect shows decreasing trend from the onset to the peak, in the rising phase.
In our analysis, if the profiles of the rising phase for different energy channels are similar, we can determine accurately $t_\eta$ for different energy channels. The onset time corresponds to $\eta \rightarrow 0$.
However contamination is most significant at small $\eta$. Therefore, to obtain a correct onset time, what one can do is obtain a series of $L_\eta$ for $\eta$ not so small and then extrapolate to $\eta$ = 0. Because the rising phase is rather rapid, deciding $t_\eta$ at $\eta$s which are not close to zero is easier. Consequently extrapolating to  $L_\eta(\eta\rightarrow 0)$ will be more accurate than the traditional VDA analysis. Therefore our analysis is less affected by contamination than the VDA analysis.

\begin{figure}[ht!]
\plotone{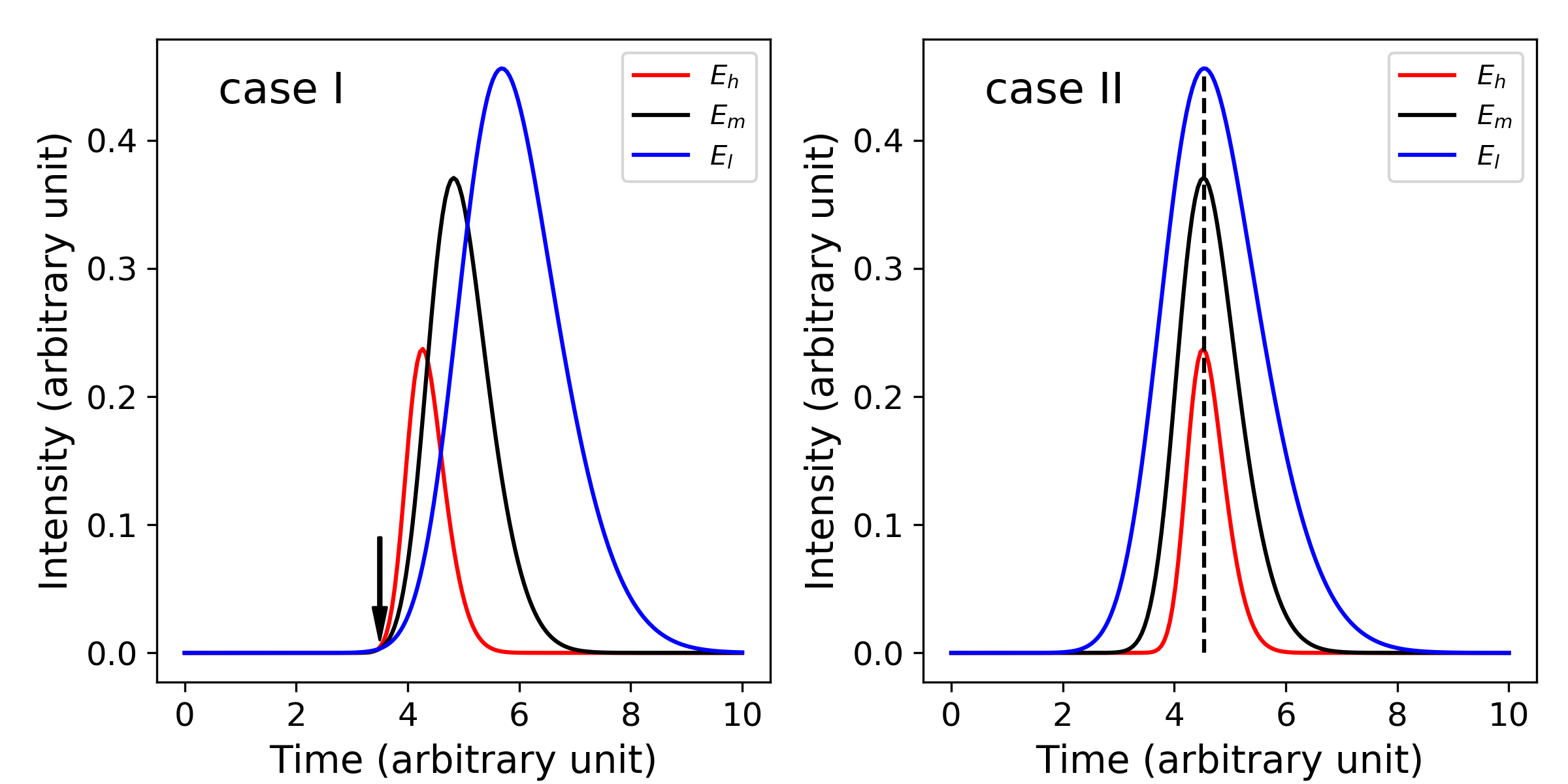}
\caption{Two onset models that we consider in this work. The left panel shows the case that electrons with different energies have the same onset time at the sun. The right panel shows the case that electrons with different energies have the same peak time at the sun. Red, black, and blue curves indicate the time intensity profiles for electrons with high, medium, and low energy.}\label{fig:onset}
\end{figure}

The most notable feature of our new method is the use of multiple $\eta$'s. It, in a way, provides a self-consistent check for the result. In the scenario that the onset times are the same for all electrons at different energies, illustrated in the left panel of figure~\ref{fig:onset}, a smaller value $\eta$ will lead to a path length closer to the real situation. 
And if the injection profile at different energies peak at the same time, as shown in the right panel in figure~\ref{fig:onset}, then we expect a larger $\eta$ gives more accurate path length. 
Indeed, by fitting the path length as a function of $\eta$ and extrapolating it to $\eta=0$ and $\eta=1$, one can get 
the corresponding path lengths for the two scenarios in figure~\ref{fig:onset}. 
If the rising phase is very short, then using different $\eta$'s may yield very close results.
As discussed above, the value of $\eta$ can not be too small in order to eliminate the energy deposition effect from the instrument. 
In this work, we use $\eta=$ 3/4, 1/2, and 1/3. 

Hereafter, we refer to our new VDA analysis as FVDA for fractional VDA.  
We note that the FVDA method is best for electrons 
and may lead to overestimate of path length for protons and ions. 
This is because the interplanetary waves and turbulence often affect more the transport of protons and ions than electrons. Indeed, by employing several assumptions about the interplanetary scattering and the injection profile, \citet{Saiz2005} examines numerically how the VDA analysis can lead to an over estimate of the path length for $2$ to $2000$ MeV protons. Since there is no background level in their numerical experiments, they tested three thresholds at $0.01\%$, $2\%$, and $60\%$ of the peak value and found that although the derived path length at the $0.01\%$ threshold is about $25\%$ larger than the true path length, it can be off by $100\%$ for the larger threshold of $60\%$. We point out that the analysis of \citet{Saiz2005} is very similar to our FVDA method except that 
their choices of the $0.01\%$ and  $60\%$ thresholds are not feasible in reality due to non-zero background. \citet{Masson2012} considered 
using a single threshold of $50\%$ the peak value for proton events with rapid rise. As shown in \citet{Saiz2005}, the derived path length can be significantly larger than the actual one.

We select events that satisfy the following criteria:
(1) the time intensity profiles of the outward propagating electrons show quick rising phases (within one hour) in a broad energy range with well-defined and similar shapes. 
(2) the event has good count statistics and the entire phase of the event can be clearly separated from the background.
With these two criteria, we obtain $125$ events.

\section{Analysis and Discussion} \label{sec:analysis}
We make no clear distinction between gradual events and impulsive events in our analysis. Gradual events with rapid onset phases and well-defined profiles which allow the determination of peak fluxes are also included in our selected events. 

The magnetic field path length is calculated by performing the linear regression of the time $t_\eta$ and the inverse of velocity $1/v_i$: 
\begin{equation} 
\frac{L_\eta}{v_i}=t_{\eta,i}-t_0.
\end{equation}
$L_\eta$ is the path length calculated using time {$t_{\eta,i}$ which gives a flux equal to $\eta (j_p-j_b)+j_b$ for energy $E_i$}, and $t_0$ is the release time at the sun. In the analysis, the velocity $v_i$ is chosen such that the corresponding $E_i$ is the center energy of the energy channels. Although the earliest arriving particles in one energy channel often have the highest energy in that energy channel, electrons in one energy channel are composed mostly by the low energy electrons (due to the negative energy spectral slope). 
To examine the effect of the choice of energy we use, we  have also performed the linear regression assuming $v_i$ to be either the lower or higher boundary of the energy channel. The results are similar, and the overall distribution discussed below does not change much.
In the following statistical discussion, we discard events that have uncertainties larger than $0.1$ AU. 
A larger uncertainty in the linear regression usually means the times, $t_{\eta,i}$, do not lie perfectly in a straight line. These could be events in which the time intensity profiles of different energy channels do not  behave similarly.
With this constraint, a total of $81$ events remain.

The events and the results are summarized in table~\ref{tbl:database}. This is our main result of the paper. The first $4$ columns contain the year, month, day, and start time of the events. The $5-7$ columns contain the calculated path length and its uncertainty when $\eta$ equals $3/4$, $1/2$, and $1/3$, respectively.
We also show the peak-to-background intensity ratio ($j_p/j_b$) in the energy channel of $82-135$ keV in column $8$, as an indicator of the relative strength of the event with respect to the pre-event background. 
This is the only common channel for all events we studied in this work. 
The time difference dT (in seconds) between the $3/4$ and $1/2$ of the peak flux is shown in column $9$, as an indicator of the duration of the rising phase. 

\begin{figure}[ht!]
\plotone{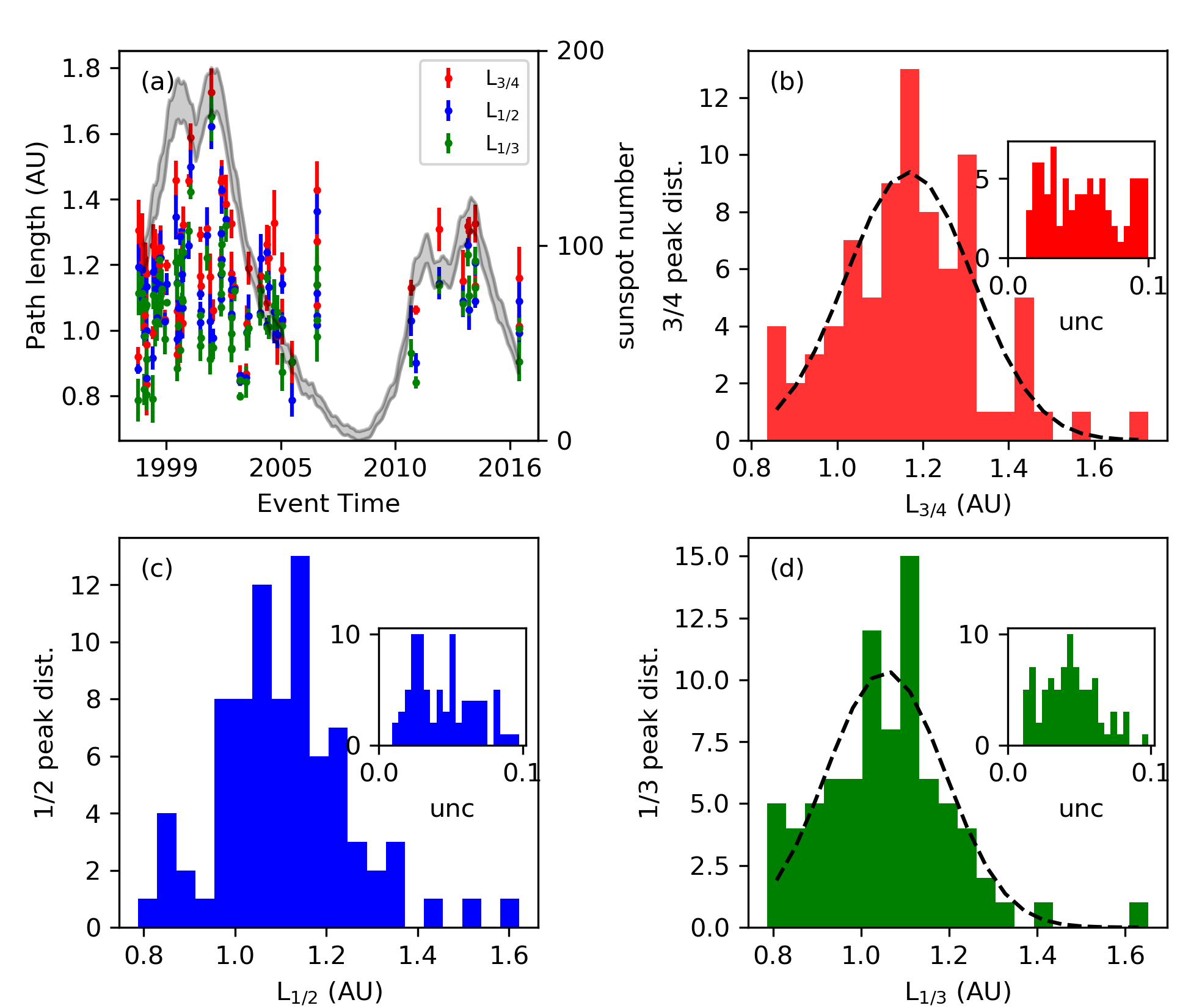}
\caption{Panel (a) plots the path lengths and their uncertainties calculated using FVDA when $\eta$ is $3/4$ (red), $1/2$ (blue), and $1/3$ (green). The shaded curve plots the monthly total sunspot number. Panel (b) plots the histogram of the path lengths and their uncertainties calculated with $\eta=3/4$. The mean and stand deviation of the path length is 1.17 AU and 0.17 AU. The black dashed curve is the fitted Gaussian distribution. 
Panel (c) plots the histogram of the path lengths and their uncertainties calculated with $\eta=1/2$. The mean and stand deviation of the path length is 1.11 AU and 0.14 AU.
Panel (d) plots the path length and their uncertainty distributions calculated with $\eta=1/3$. The mean and stand deviation of the path length is 1.06 AU and 0.15 AU.
The black dashed curve is fitted Gaussian distribution.}\label{fig:dist}
\end{figure}

Panel (a) in figure~\ref{fig:dist} is a scattered plot of the calculated path lengths with uncertainties for all events. The red, blue, and green color represent the path length calculated with $\eta$ equals $3/4$, $1/2$, and $1/3$.
The grey shaded curve plot the $13$-month smoothed monthly total sunspot number obtained from \url{http://www.sidc.be/silso/datafiles}.
The distribution of the path lengths do not show clear correlation with respect to the sunspot number. 
The histograms of the path lengths and their uncertainties are plotted in panels (b), (c), and (d). 
The corresponding mean and standard deviation is $1.17$ AU and $0.17$ AU for $\eta=3/4$, $1.11$ AU and $0.14$ AU for $\eta=1/2$, and $1.06$ AU and $0.15$ AU for $\eta=1/3$. One may also try to fit the histogram with a Gaussian distribution. For panel (b) and (d), the 
Gaussian fits are plotted by the black dashed-line (for panel (c), the Shapiro-Wilk test \citep{Shapiro1965}
rejects the Gaussian hypothesis of the histogram at $\alpha$-value of $0.05$.)
We note that as $\eta$ gets smaller, the mean path length shifts to a smaller values. For any given event, choosing a smaller $\eta$ does not necessarily lead to a more accurate result of path length. As explained before, 
only when the release time at different energies are the same, choosing a smaller $\eta$ yields a better estimation of the path length. However, as mentioned above, the contamination effect from the higher energy channel is stronger when $\eta$ is smaller. This contamination will lead to earlier elevations of the low energy channels. Consequently, the calculated path length will be smaller than the actual one. Because contamination is more pronounced at smaller $\eta$, one expects that as $\eta$ becomes smaller, the path length also decreases. Indeed, this is what we found 
from our analysis. This also explains to some extend
 the fact that there are $11$, $17$, and $26$ (out of $81$) events which have calculated path lengths less than $1$ AU for $\eta=3/4$, $1/2$, and $1/3$, respectively. 
There is another reason why we obtain path lengths less than 1 AU: if low energy electrons are released earlier than high energy electrons, our analysis will also yield a shorter path length than the actual value. One can turn this argument around and argue that the events whose path length are smaller than $1$ AU must not have the same release times for electrons at different energies. In any case, our analysis suggests that using multiple $\eta$ will allow us to obtain some constraints on the release profiles at the sun. 


Although we are focusing on the rising phases of these events, electrons however undergo pitch angle scattering as they propagate out to $1$ AU due to interplanetary turbulence. \citet{Tan2011,Tan2013} suggested that the first arriving electrons may not be scatter-free due to pre-existing R-mode and L-mode waves which can scatter non-relativistic electrons.
Moreover, a transition from the scatter-free to diffusive propagation of electrons was reported by \citet{Tan2011}. 
The transition energy is found to be event dependent and it is between $\sim 60$ keV and $\sim 120$ keV in one event and between $\sim 250$ keV and $\sim 500$ keV in another.
Nevertheless, there are only $2$, $1$, $1$ events 
(for $\eta=3/4$, $1/2$, and $1/3$ respectively) that have calculated path length greater than $1.5$ AU.
Earlier, \citet{Kahler2006} calculated the travel distance of $80$ electrons detected by WIND SST instrument and found the path lengths ranged from $0.15$ to $2.7$ AU. However, even in their analysis, there were only two events that have path length greater than $2.0$ AU; and the majority of their events have path lengths shorter than $1.5$ AU.

\begin{figure}[ht!]
\plotone{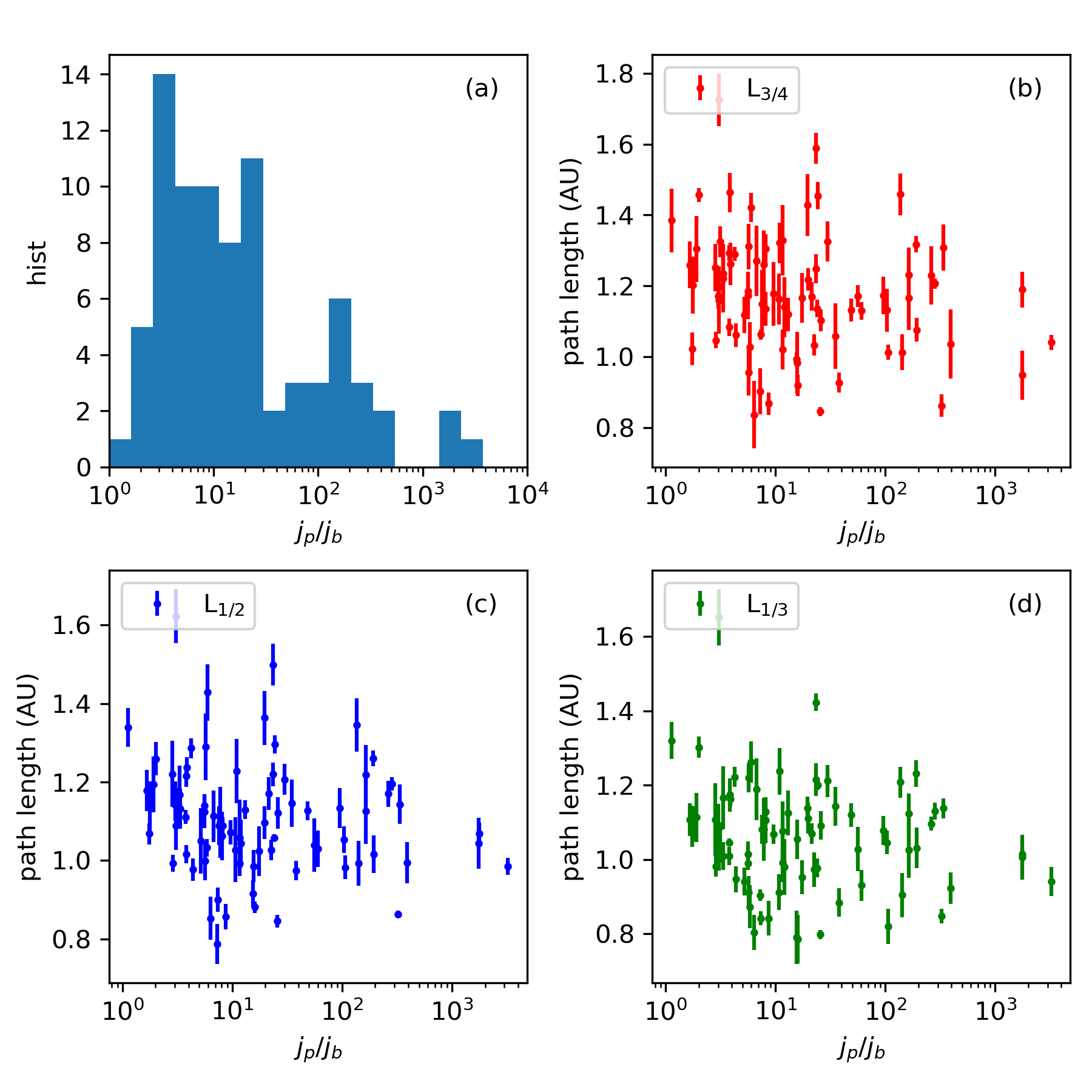}
\caption{Panel (a) plots the histogram of the peak-to-background intensity ratio ($j_p/j_b$) in the energy range of $82-135$ keV. Panels (b), (c), and (d) plot the calculated path length with respect to the peak-to-background ratio when $\eta$ is $3/4$, $1/2$, and $1/3$.}\label{fig:peakdist}
\end{figure}

Figure~\ref{fig:peakdist} shows the dependence of the calculated path length on the peak-to-background ratio ($j_p$/$j_b$) of the $82-135$ keV electrons. Panel (a) shows the histogram of $j_p$/$j_b$ and panel (b), (c), and (d) plot the calculated path length and its uncertainties with respect to $j_p$/$j_b$.
The plots do not show a clear correlation between the path length and the $j_p$/$j_b$ ratio, indicating that our method is not biased toward larger or smaller events. This is different from previous work of 
\citet{Kahler2006}, who compared the path length with the corresponding $82$ keV peak-to-background ratios and found a weak correlation. In the simulations of \citet{Lintunen2004}, an association of longer path length with smaller peak intensities is obtained because of the delayed onset times of low-energy channels compared to the high-energy channels. And more intense events are associated with slightly smaller path length.
These are because the determination of the onset time is affected by the background count level as well as the spectral shape.
To determine the onset time, one often uses the criterion that the intensity exceeds a fixed fraction $k$ (from $0.001$ to $0.1$) of the peak intensity.
In the work of \citet{Lintunen2004}, the authors found that increasing the fraction, $k$, by an order of magnitude will increase the path length by $30-50\%$. 

\begin{figure}[ht!]
\plotone{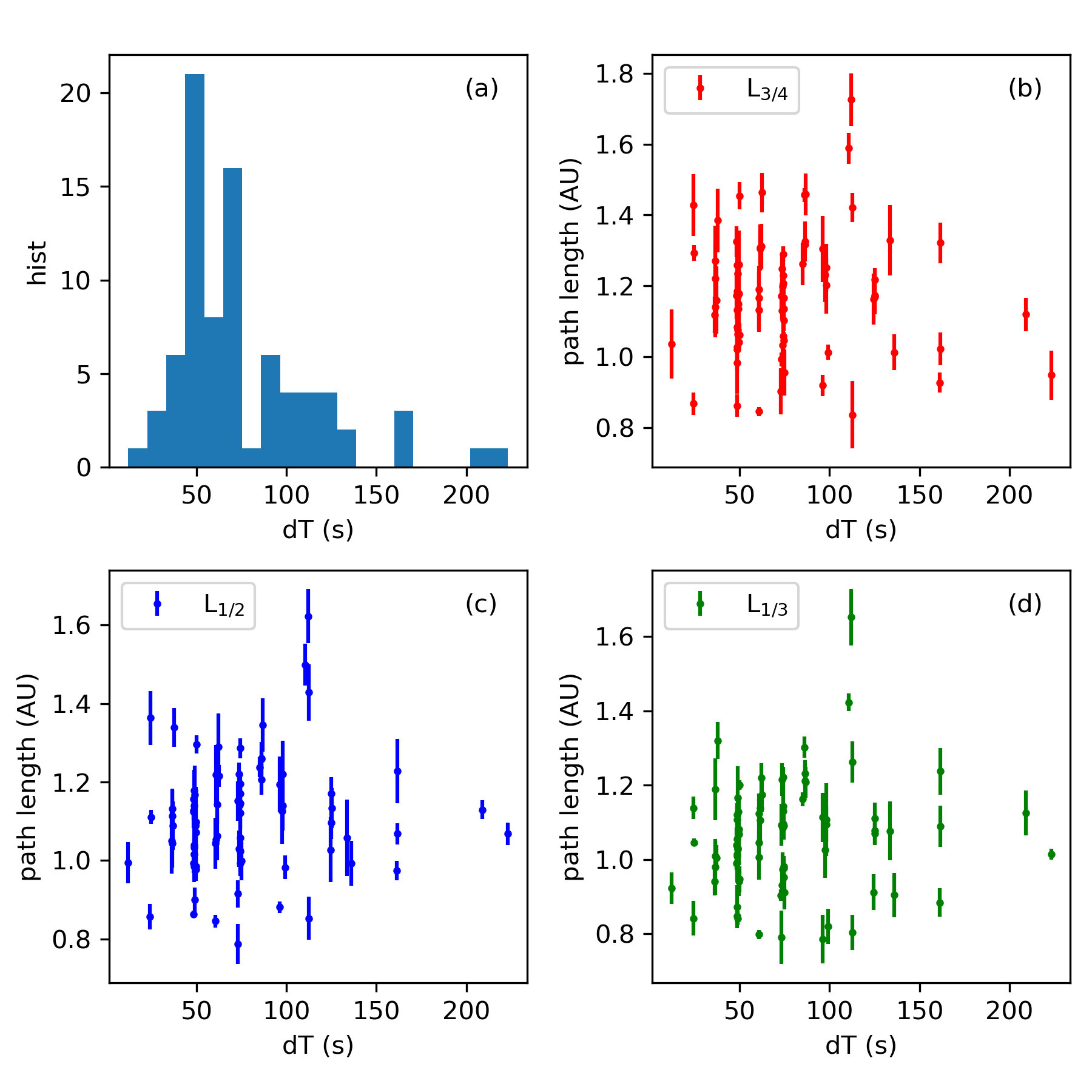}
\caption{Panel (a) plots the histogram of the onset time scale in the energy range of $82-135$ keV. Panels (b), (c), and (d) plot the calculated path length with respect to the onset time scale when $\eta$ is $3/4$, $1/2$, and $1/3$.\label{fig:dtdist}}
\end{figure}

Figure~\ref{fig:dtdist} shows the dependence of the calculated path length on the duration of the rising phase of the event. We use the time difference between the $3/4$ peak time and the $1/2$ peak time as a proxy for the 
duration of the rising phase, and in the following we 
refer this as the onset time scale. Panel (a) plots the histogram of the onset time scale of all events and panels (b), (c), and (d) plot the correlation between the onset time scale and the calculated path length for the three choices of $\eta=$ $3/4$, $1/2$, and $1/3$.  There is no clear correlation between the path length and the onset time scale. This suggests that one can extend the FVDA method to events where the rising phase does not need to be very sharp. 

\section{Conclusion}\label{sec:conclusion}
Understanding the configuration of interplanetary magnetic field is important to understand the transport of energetic particles in SEP events. In particular, the configuration of interplanetary magnetic field can put strong constraints on particles' cross field diffusion. Previous studies have obtained values of $\kappa_{\perp}/\kappa_{||}$ which differ considerably for the dropout events and the wide-spreading events \citep{Mazur2000, Giacalone2000, Dresing2014}. To reconcile this discrepancy, one possible explanation is that magnetic field configurations can differ considerably in different events. For example, if there is significant field line meandering in the interplanetary medium \citep{Laitinen2016}, the same event can be seen from multiple spacecraft which have large longitudinal separation. The field line meandering effect may also vary with different interplanetary turbulence conditions.

In this work, we calculate the interplanetary magnetic path length using a newly developed method: the fraction velocity dispersion analysis (FVDA) method. This method does not require an accurate determination of the onset time of electrons as in the standard VDA method. It is therefore less affected by the background flux. It also does not require an accurate determination of the 
peak time, as done in \citet{Wang2016}. Instead of considering either the peak time or the onset time, the FVDA utilizes the times in the rising phase of an event that correspond to the flux that is a fraction, $\eta$, of the peak flux. We applied the FVDA method to electron events that have a well-defined peak and relative prompt onset phase. Note that a stronger scattering usually leads to a prolonged onset phase. Therefore, our selection criteria naturally eliminate very diffusive events and events with extended injection. 
Using this method, we identified $81$ electron events of which path lengths are obtained with uncertainties less than $0.1$ AU. And the calculated path lengths using different $\eta$s yield similar results.
The obtained path lengths in these events are very close to the nominal Parker field lengths, suggesting that the magnetic field themselves may also be close to the Parker configuration. 
From the distribution of the path length with respect to the peak-to-background ratios and the onset time scales, our method does not show bias toward either large or small, fast or slow onset events. The mean and stand deviation of the calculated path lengths are ($1.17$ AU, $0.17$ AU) for $\eta=3/4$, ($1.11$ AU, $0.14$ AU) for $\eta=1/2$, and ($1.06$ AU, $0.15$ AU) for $\eta=1/3$. And the distribution of the calculated path lengths for $\eta=3/4$ and $1/3$ are well represented by a Gaussian distribution.

The FVDA method we developed here is an extension of VDA. Comparing to the traditional VDA analysis, it is less affected by the uncertainty in determining the onset time or the peak time of the time intensity profiles. However, because it makes use of the rapid rising phase, so if the rising phases for different energy channels are not similar, the resulting path length from FVDA can be also nonphysical. This puts some constraints in applying the FVDA method. In our current work, we find that the FVDA applies nicely to "nearly-scatter-free" electron events. We caution that one has to take extreme care when applying FVDA to other events such as ion events or events that are scatter-dominated.

For a turbulent magnetic field, \citet{Ragot2008} suggested that a correction factor should be applied to the length of the smooth magnetic field length. And the correction factor is found to be $1.16\pm0.06$ and $1.23\pm0.03$ in slow and fast solar wind at $0.3$ AU, and $1.45\pm0.25$ and $1.33\pm0.06$ in slow and fast solar wind at $1$ AU. This is not what we find in this work. Our results of the magnetic field lengths indicate that the interplanetary magnetic field paths traveled by the $27-310$ keV electrons are close to the ideal Parker length. One may expect that the path lengths are correlated with solar activity. However, the events in our study span over a full solar cycle and show no clear dependence on solar cycle. Therefore our results suggest that the interplanetary magnetic field does not differ much from the Parker's spiral field during both solar maximum or solar minimum. 
This is somewhat counter intuitive in that we expect the solar wind magnetic field is affected more during solar maximum. However, we note that in our analysis we required the background to be reasonably quiet even in solar maximum: no large preceding events occurred in our selection. Although there was no large preceding events, there still could be many small (or nano) eruptions. One may expect that these small eruptions can also change the interplanetary magnetic field configurations. Our results, however, suggest that they do not. 
Note that the change of the Parker spiral length from $1.14$ AU (solar wind speed of $\sim 400$ km/s) to 1.03 AU (solar wind speed of $800$ km/s) is comparable to the uncertainties of the path length calculated in this work. Our results has important implications for the study of energetic particle transport in the solar wind. Further detailed studies for selected events in table~\ref{tbl:database} will be reported in a separate paper. 

Acknowledgement: L.Z. and M.Z. are supported at Florida Institute of Technology under NNX15AN72G, NNX15AB76G, 80NSSC19K0076, and 80NSSC18K0644; G.L. and A.M. are supported at University of Alabama in Huntsville under NNX17AI17G, NNX17AK25G, and 80NSSC19K0075. L.W. thanks NSFC for support under grants 41774183 and 41861134033. F.E. is supported by NNX17AK25G at the Bay Area Environmental Research Institute. {G.L. and F.E. also acknowledge supports from the International Space Science Institute (ISSI) through the team on 'Solar flare acceleration signatures and their connection to solar energetic particles'. In particular discussions with Drs. L. Klein, T. Laitinen, N. Bian, and Du Toit Strauss.}

\begin{center}
\begin{center}
 \begin{longtable}{cccccccccc}
  \caption{Electron Event Data}\\
    \toprule
    \toprule
    \multicolumn{1}{c}{year} & \multicolumn{1}{c}{month} & 
    \multicolumn{1}{c}{day} &  \multicolumn{1}{c}{hour} & 
    \multicolumn{1}{c}{\rm{$L_{3/4}$}} &   \multicolumn{1}{c}{\textbf{$L_{1/2}$}}  &
    \multicolumn{1}{c}{$L_{1/3}$} &   \multicolumn{1}{c}{$j_p$/$j_b$} & 
    \multicolumn{1}{c}{dT} \\
 
 \multicolumn{1}{c}{\textbf{}} & \multicolumn{1}{c}{\textbf{}} & \multicolumn{1}{c}{} &
 \multicolumn{1}{c}{(h)} & \multicolumn{1}{c}{(AU)} & \multicolumn{1}{c}{(AU)} & 
 \multicolumn{1}{c}{(AU)} & \multicolumn{1}{c}{} & 
 \multicolumn{1}{c}{(s)} \\
    \midrule
    \endhead
    1998  & 5     & 2     & 4.7   & 0.92$\pm$0.03 & 0.88$\pm$0.01 & 0.79$\pm$0.07 & 15.93 & 48.07 \\
    1998  & 5     & 8     & 2.0   & 1.3$\pm$0.09 & 1.19$\pm$0.07 & 1.11$\pm$0.07 & 1.92  & 48.13 \\
    1998  & 7     & 11    & 22.73 & 1.26$\pm$0.1 & 1.1$\pm$0.09 & 1.08$\pm$0.08 & 7.8   & 49.59 \\
    1998  & 7     & 11    & 23.5  & 1.2$\pm$0.02 & 1.19$\pm$0.02 & 1.11$\pm$0.02 & 282.57 & 49.59 \\
    1998  & 8     & 13    & 15.0  & 1.01$\pm$0.02 & 0.98$\pm$0.03 & 0.82$\pm$0.05 & 106.27 & 74.43 \\
    1998  & 8     & 13    & 17.78 & 1.1$\pm$0.03 & 1.12$\pm$0.04 & 1.09$\pm$0.04 & 25.82 & 74.43 \\
    1998  & 8     & 29    & 18.45 & 1.05$\pm$0.02 & 0.99$\pm$0.02 & 0.98$\pm$0.03 & 2.86  & 49.63 \\
    1998  & 9     & 8     & 15.25 & 1.18$\pm$0.09 & 1.07$\pm$0.03 & 1.07$\pm$0.03 & 9.6   & 24.81 \\
    1998  & 9     & 25    & 22.25 & 0.96$\pm$0.06 & 1.0$\pm$0.05 & 0.91$\pm$0.05 & 5.68  & 50.03 \\
    1998  & 9     & 26    & 16.25 & 0.84$\pm$0.1 & 0.85$\pm$0.05 & 0.8$\pm$0.05 & 6.37  & 37.52 \\
    1998  & 9     & 27    & 8.08  & 1.17$\pm$0.05 & 1.13$\pm$0.05 & 1.08$\pm$0.04 & 95.5  & 100.05 \\
    1999  & 1     & 7     & 0.2   & 0.99$\pm$0.02 & 0.92$\pm$0.03 & 0.79$\pm$0.07 & 15.48 & 48.76 \\
    1999  & 1     & 24    & 16.23 & 1.26$\pm$0.07 & 1.18$\pm$0.05 & 1.11$\pm$0.04 & 1.66  & 48.76 \\
    1999  & 2     & 20    & 3.9   & 1.23$\pm$0.08 & 1.13$\pm$0.08 & 1.03$\pm$0.07 & 164.5 & 48.77 \\
    1999  & 2     & 20    & 20.52 & 1.17$\pm$0.04 & 1.15$\pm$0.05 & 1.09$\pm$0.06 & 3.03  & 48.77 \\
    1999  & 3     & 21    & 16.83 & 1.17$\pm$0.03 & 1.04$\pm$0.07 & 1.03$\pm$0.06 & 56.07 & 48.87 \\
    1999  & 5     & 8     & 14.5  & 1.2$\pm$0.08 & 1.14$\pm$0.06 & 1.1$\pm$0.05 & 1.77  & 24.5 \\
    1999  & 5     & 12    & 5.42  & 1.23$\pm$0.06 & 1.17$\pm$0.07 & 1.17$\pm$0.08 & 3.33  & 49.02 \\
    1999  & 5     & 12    & 6.73  & 1.25$\pm$0.04 & 1.22$\pm$0.03 & 1.22$\pm$0.04 & 23.48 & 24.51 \\
    1999  & 5     & 27    & 21.85 & 1.25$\pm$0.07 & 1.22$\pm$0.08 & 1.11$\pm$0.1 & 2.83  & 49.02 \\
    1999  & 6     & 18    & 11.51 & 1.12$\pm$0.05 & 1.13$\pm$0.02 & 1.13$\pm$0.06 & 13.04 & 135.2 \\
    1999  & 8     & 7     & 17.0  & 1.03$\pm$0.03 & 1.03$\pm$0.03 & 0.97$\pm$0.05 & 22.58 & 24.65 \\
    1999  & 9     & 11    & 2.82  & 1.2$\pm$0.02 & 1.14$\pm$0.03 & 1.08$\pm$0.01 & 3.05  & 24.67 \\
    2000  & 2     & 18    & 9.22  & 1.46$\pm$0.06 & 1.35$\pm$0.07 & 1.21$\pm$0.04 & 136.39 & 74.34 \\
    2000  & 3     & 7     & 7.37  & 0.93$\pm$0.03 & 0.97$\pm$0.02 & 0.88$\pm$0.04 & 38.05 & 37.18 \\
    2000  & 3     & 7     & 12.42 & 1.06$\pm$0.09 & 1.15$\pm$0.06 & 1.14$\pm$0.05 & 34.94 & 74.37 \\
    2000  & 4     & 4     & 15.17 & 0.95$\pm$0.07 & 1.07$\pm$0.03 & 1.01$\pm$0.01 & 1771.21 & 123.94 \\
    2000  & 5     & 1     & 10.13 & 1.04$\pm$0.02 & 0.99$\pm$0.02 & 0.94$\pm$0.04 & 3237.1 & 49.6 \\
    2000  & 5     & 1     & 13.0  & 1.29$\pm$0.02 & 1.29$\pm$0.03 & 1.22$\pm$0.03 & 4.23  & 74.4 \\
    2000  & 6     & 4     & 6.92  & 1.23$\pm$0.08 & 1.17$\pm$0.03 & 1.1$\pm$0.02 & 262.84 & 37.2 \\
    2000  & 6     & 16    & 2.83  & 1.02$\pm$0.05 & 1.07$\pm$0.03 & 1.09$\pm$0.05 & 1.76  & 0.0 \\
    2000  & 6     & 23    & 14.35 & 1.32$\pm$0.06 & 1.23$\pm$0.08 & 1.24$\pm$0.06 & 10.96 & 74.54 \\
    2000  & 9     & 30    & 22.41 & 1.46$\pm$0.02 & 1.26$\pm$0.04 & 1.3$\pm$0.03 & 2.01  & 73.7 \\
    2000  & 10    & 30    & 2.95  & 1.59$\pm$0.04 & 1.5$\pm$0.05 & 1.42$\pm$0.02 & 23.51 & 85.98 \\
    2001  & 4     & 24    & 19.95 & 1.29$\pm$0.02 & 1.11$\pm$0.02 & 1.05$\pm$0.01 & 3.78  & 24.83 \\
    2001  & 4     & 25    & 13.72 & 1.17$\pm$0.07 & 1.02$\pm$0.06 & 0.95$\pm$0.05 & 17.41 & 62.08 \\
    2001  & 4     & 30    & 10.77 & 1.14$\pm$0.02 & 1.06$\pm$0.01 & 0.98$\pm$0.03 & 23.98 & 62.08 \\
    2001  & 8     & 13    & 18.35 & 1.31$\pm$0.06 & 1.29$\pm$0.08 & 1.22$\pm$0.04 & 5.71  & 49.68 \\
    2001  & 10    & 9     & 7.51  & 1.16$\pm$0.07 & 1.03$\pm$0.08 & 0.91$\pm$0.05 & 10.7  & 74.65 \\
    2001  & 10    & 28    & 11.07 & 1.73$\pm$0.07 & 1.62$\pm$0.07 & 1.65$\pm$0.08 & 3.06  & 62.21 \\
    2001  & 12    & 4     & 7.9   & 1.06$\pm$0.03 & 0.98$\pm$0.03 & 0.95$\pm$0.03 & 4.36  & 49.89 \\
    2002  & 4     & 14    & 22.32 & 1.17$\pm$0.04 & 1.17$\pm$0.04 & 1.07$\pm$0.03 & 21.42 & 62.5 \\
    2002  & 4     & 15    & 2.7   & 1.45$\pm$0.04 & 1.3$\pm$0.02 & 1.2$\pm$0.01 & 24.37 & 37.5 \\
    2002  & 4     & 15    & 17.48 & 1.22$\pm$0.03 & 1.1$\pm$0.04 & 1.11$\pm$0.04 & 19.76 & 62.5 \\
    2002  & 4     & 25    & 5.85  & 1.42$\pm$0.04 & 1.43$\pm$0.07 & 1.26$\pm$0.05 & 5.96  & 62.5 \\
    2002  & 5     & 1     & 19.15 & 1.46$\pm$0.06 & 1.22$\pm$0.03 & 1.17$\pm$0.04 & 3.82  & 50.0 \\
    2002  & 7     & 18    & 23.92 & 1.38$\pm$0.09 & 1.34$\pm$0.05 & 1.32$\pm$0.05 & 1.13  & 37.69 \\
    2002  & 10    & 19    & 21.13 & 1.17$\pm$0.07 & 1.12$\pm$0.04 & 0.99$\pm$0.02 & 5.6   & 24.19 \\
    2002  & 10    & 20    & 11.52 & 1.33$\pm$0.04 & 1.16$\pm$0.03 & 1.04$\pm$0.05 & 3.12  & 12.1 \\
    2002  & 10    & 20    & 14.05 & 1.11$\pm$0.1 & 1.04$\pm$0.04 & 0.95$\pm$0.04 & 391.81 & 24.19 \\
    2002  & 10    & 21    & 4.2   & 1.12$\pm$0.05 & 1.05$\pm$0.08 & 0.94$\pm$0.04 & 5.19  & 24.19 \\
    2002  & 12    & 12    & 12.55 & 1.13$\pm$0.03 & 1.13$\pm$0.02 & 1.12$\pm$0.03 & 48.6  & 48.48 \\
    2003  & 3     & 17    & 10.05 & 0.86$\pm$0.03 & 0.86$\pm$0.01 & 0.85$\pm$0.02 & 323.42 & 36.4 \\
    2003  & 3     & 17    & 16.48 & 0.84$\pm$0.01 & 0.85$\pm$0.02 & 0.8$\pm$0.01 & 25.49 & 24.26 \\
    2003  & 6     & 29    & 8.03  & 0.87$\pm$0.03 & 0.86$\pm$0.03 & 0.84$\pm$0.05 & 8.66  & 24.26 \\
    2003  & 7     & 8     & 2.3   & 1.02$\pm$0.06 & 0.99$\pm$0.03 & 0.99$\pm$0.05 & 11.65 & 24.26 \\
    2003  & 8     & 10    & 19.4  & 1.19$\pm$0.05 & 1.04$\pm$0.07 & 1.01$\pm$0.06 & 1759.73 & 36.4 \\
    2004  & 2     & 28    & 3.23  & 1.13$\pm$0.06 & 1.05$\pm$0.03 & 1.05$\pm$0.03 & 103.53 & 36.44 \\
    2004  & 3     & 16    & 8.92  & 1.17$\pm$0.09 & 1.22$\pm$0.08 & 1.12$\pm$0.05 & 163.61 & 145.77 \\
    2004  & 6     & 25    & 16.5  & 1.08$\pm$0.03 & 1.02$\pm$0.02 & 1.01$\pm$0.02 & 3.79  & 48.59 \\
    2004  & 6     & 27    & 5.03  & 1.26$\pm$0.06 & 1.24$\pm$0.03 & 1.16$\pm$0.02 & 3.88  & 36.45 \\
    2004  & 7     & 24    & 18.6  & 1.22$\pm$0.1 & 1.13$\pm$0.05 & 1.01$\pm$0.04 & 3.34  & 12.15 \\
    2004  & 11    & 1     & 5.88  & 1.33$\pm$0.1 & 1.06$\pm$0.1 & 1.08$\pm$0.08 & 11.67 & 158.01 \\
    2004  & 12    & 25    & 22.35 & 0.98$\pm$0.09 & 0.99$\pm$0.04 & 1.05$\pm$0.05 & 15.68 & 36.47 \\
    2005  & 3     & 16    & 19.75 & 1.03$\pm$0.07 & 1.03$\pm$0.02 & 0.87$\pm$0.06 & 5.88  & 85.09 \\
    2005  & 3     & 16    & 23.0  & 1.19$\pm$0.05 & 1.14$\pm$0.03 & 1.01$\pm$0.03 & 5.61  & 36.47 \\
    2005  & 9     & 4     & 14.95 & 0.9$\pm$0.06 & 0.79$\pm$0.05 & 0.9$\pm$0.02 & 7.26  & 72.96 \\
    2006  & 11    & 19    & 22.86 & 1.43$\pm$0.09 & 1.36$\pm$0.07 & 1.14$\pm$0.03 & 19.59 & 36.62 \\
    2006  & 11    & 20    & 3.42  & 1.08$\pm$0.03 & 1.02$\pm$0.05 & 1.03$\pm$0.05 & 193.99 & 36.62 \\
    2006  & 11    & 21    & 8.32  & 1.14$\pm$0.08 & 1.04$\pm$0.06 & 0.98$\pm$0.08 & 12.03 & 24.41 \\
    2006  & 11    & 21    & 16.63 & 1.27$\pm$0.1 & 1.11$\pm$0.07 & 1.19$\pm$0.08 & 6.75  & 36.62 \\
    2011  & 5     & 15    & 23.44 & 1.13$\pm$0.02 & 1.03$\pm$0.05 & 0.93$\pm$0.04 & 60.4  & 36.79 \\
    2011  & 8     & 8     & 15.86 & 1.06$\pm$0.01 & 0.9$\pm$0.03 & 0.84$\pm$0.02 & 7.39  & 36.79 \\
    2012  & 9     & 19    & 4.54  & 1.31$\pm$0.07 & 1.14$\pm$0.05 & 1.14$\pm$0.03 & 336.08 & 49.1 \\
    2013  & 11    & 12    & 8.12  & 1.15$\pm$0.1 & 1.09$\pm$0.05 & 1.08$\pm$0.04 & 7.52  & 61.55 \\
    2014  & 2     & 5     & 22.92 & 1.32$\pm$0.02 & 1.26$\pm$0.02 & 1.23$\pm$0.04 & 191.74 & 36.95 \\
    2014  & 2     & 20    & 7.74  & 1.3$\pm$0.04 & 1.06$\pm$0.06 & 1.11$\pm$0.06 & 8.11  & 36.95 \\
    2014  & 6     & 12    & 10.8  & 1.14$\pm$0.05 & 1.09$\pm$0.02 & 1.13$\pm$0.03 & 8.18  & 86.25 \\
    2014  & 6     & 12    & 13.22 & 1.33$\pm$0.06 & 1.21$\pm$0.04 & 1.21$\pm$0.04 & 29.75 & 49.28 \\
    2016  & 7     & 20    & 22.0  & 1.01$\pm$0.05 & 0.99$\pm$0.06 & 0.9$\pm$0.06 & 141.9 & 37.07 \\
    2016  & 7     & 21    & 7.0   & 1.16$\pm$0.09 & 1.09$\pm$0.06 & 1.0$\pm$0.03 & 3.07  & 24.72 \\
      \label{tbl:database}
\end{longtable}
\end{center}

\end{center}
\clearpage

\clearpage

\clearpage

\clearpage

\clearpage

\end{document}